\documentclass[,final            
  ,numberedheadings 
  ]
  {aipproc}
\usepackage{amsmath,supertabular}
\usepackage{amssymb,mathrsfs}
\layoutstyle{6x9}
\def\set#1{{\sf #1}}

\def\d{\operatorname{d}}\def\<{\langle}\def\>{\rangle}
\def\Tr{\operatorname{Tr}}\def\:{\hbox{\bf :}}

\def\St#1{\set{St}(#1)}
\def\Lin#1{\set{Lin} \left( #1 \right)}
\def\Detcomb#1{\set{DetComb} \left( #1 \right)}
\def\Probcomb#1{\set{ProbComb} \left( #1 \right)}

\def\map#1{{\mathscr{#1}}}

\def\set#1{{\sf #1}}
\def\grp#1{{\mathbf #1}}

\def\Supp{\set{Supp}}\def\Span{\set{Span}}

\def\qed{$\,\blacksquare$\par}
\def\SU#1{\mathbb{SU}(#1)}

\def\dag{\dagger}
\def\spc#1{\mathscr{#1}}
 \def\kk{\rangle\!\rangle}\def\bb{\langle\!\langle}

\newtheorem{Prop}{Proposition}

\newtheorem{Theo}{Theorem}
\def\Proof{\medskip\par\noindent{\bf Proof. }}

\keywords{}\pacs{}
\begin{document}
\title{Optimal covariant quantum networks}
\author{Giulio Chiribella, Giacomo Mauro D'Ariano, Paolo Perinotti}{address ={{\em QUIT} Group,    Dipartimento di Fisica ``A. Volta'', via Bassi 6,  27100 Pavia, Italy}}

\begin{abstract} 
  A sequential network of quantum operations is efficiently described by its
  \emph{quantum comb} \cite{combs}, a non-negative operator with
  suitable normalization constraints. Here we analyze the case of
  networks enjoying symmetry with respect to the action of a given group
  of physical transformations, introducing the notion of covariant combs
  and  testers, and proving the basic structure theorems for these objects.  As an
  application, we discuss the optimal alignment of reference frames (without pre-established common references)
  with multiple rounds of quantum communication, showing that
  $\emph{i)}$ allowing an arbitrary amount of classical communication does not improve the alignment,
  and \emph{ii)} a single round of quantum communication is sufficient.
\end{abstract}
\maketitle

A quantum comb \cite{combs} describes a quantum network with $N$ open
slots in which an ordered sequence of variable quantum devices can be
inserted, thus programming the quantum operation of the resulting
circuit. Mathematically, a comb implements an \emph{admissible
  supermap} \cite{supermaps,comblong}, that transforms an input
network of $N$ quantum operations into an output quantum operation.
Having at disposal a suitable formalism opens the possibility of
optimizing the architecture of quantum circuits for a large number of
computational, cryptographic, and game-theoretical tasks, such as
discrimination of single-party strategies, cloning of quantum
transformations, and storing of quantum algorithms into quantum
memories \cite{combs,watgut,memorydisc,clonunit}. For example, quantum
combs allow one to find the optimal networks for the estimation of an
unknown group transformation with $N$ uses at disposal, a problem that
has been solved in the past only in the particular case of phase
estimation \cite{mosca}.  Using combs and supermaps one can prove in
full generality that a parallel disposition of the $N$ black boxes is
sufficient to achieve the optimal estimation of the unknown group
element \cite{memorydisc}, thus reducing the problem to the optimal
parallel estimation of group transformations already solved in Ref.
\cite{EntEstimation}.

In this paper we summarize the main concepts and methods developed so
far in the optimization of quantum networks, with focus on the case of
networks with symmetry properties, and we present an original result
on multi-round protocols for reference frame alignment.

\section{Basic notions of quantum circuits architecture}  
\subsection{Quantum $N$-combs} 
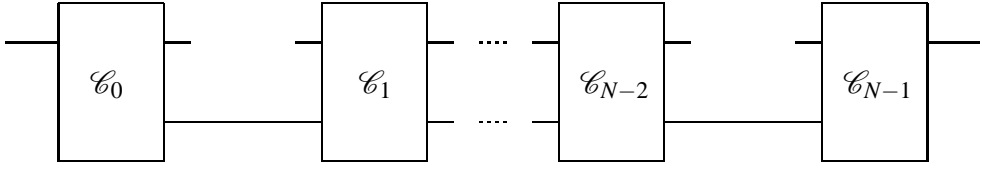
\begin{figure}[h]
\setlength{\unitlength}{.7cm}
\begin{picture}(19,5.5)(0,0) 
  \put(0,2.75){\line(1,0){1}}
  \put(3,2.75){\line(1,0){.5}}
  \put(5.5,2.75){\line(1,0){.5}}
  \put(8,2.75){\line(1,0){.5}}
  \put(10,2.75){\line(1,0){.5}}
  \put(12.5,2.75){\line(1,0){.5}}
  \put(15,2.75){\line(1,0){.5}}
  \put(17.5,2.75){\line(1,0){1}}
  \put(1,0.5){\line(0,1){3}}
  \put(3,0.5){\line(0,1){3}}
  \put(1,0.5){\line(1,0){2}}
  \put(1,3.5){\line(1,0){2}}
  \put(6,0.5){\line(0,1){3}}
  \put(8,0.5){\line(0,1){3}}
  \put(6,0.5){\line(1,0){2}}
  \put(6,3.5){\line(1,0){2}}
  \put(15.5,0.5){\line(0,1){3}}
  \put(17.5,0.5){\line(0,1){3}}
  \put(15.5,0.5){\line(1,0){2}}
  \put(15.5,3.5){\line(1,0){2}}
  \put(10.5,.5){\line(0,1){3}}
  \put(12.5,.5){\line(0,1){3}}
  \put(10.5,.5){\line(1,0){2}}
  \put(10.5,3.5){\line(1,0){2}}
  \put(3,1.25){\line(1,0){3}}
  \put(8,1.25){\line(1,0){.5}}
  \put(10,1.25){\line(1,0){.5}}
  \put(12.5,1.25){\line(1,0){3}}

  \put(1.6,1.8){$\map C_0$}
  \put(6.7,1.8){$\map C_1$}
  \put(10.9,1.8){$\map C_{N-2}$}
  \put(15.9,1.8){$\map C_{N-1}$}
 
  \multiput(9,2.75)(0.15,0){4}
  {\line(1,0){0.05}}
  \multiput(9,1.25)(0.15,0){4}
  {\line(1,0){0.05}}
\end{picture}
\caption{\label{memch} $N$-comb: sequential network of $N$ quantum operations with memory. The network contains input and output systems (free wires in the diagram), as well as internal memories (wires connecting the boxes).}
\end{figure}

Consider a sequential network of $N$
quantum operations (QOs) with memory, as in Fig.  \ref{memch}.  Due to the
presence of internal memories, there can be other networks that are
indistinguishable from it in all experiments that involve only the
incoming and outgoing quantum systems.  The quantum comb is the
equivalence class of all  networks having the same input/output
relations, irrespectively of what happens inside.
The equivalence class is in one-to-one correspondence with the Choi
operator of the network, which can be computed as the \emph{link
  product} \cite{combs} of the Choi operators of the QOs $(\map C_i)_{i=0}^{N-1}$. Here we adopt the convention that
the input (output) spaces for the QO $\map C_j$ are
indicated as $\spc H_{2j}~ (\spc H_{2j+1})$. Accordingly, the Choi
operator of the network is a non-negative operator $R^{(N)} \in
\Lin{\bigotimes_{j=0}^{2N-1} \spc H_j}$. The quantum comb can be then identified with such a Choi operator.   For networks of channels
(trace-preserving QOs) one has the recursive normalization condition
\begin{equation}\label{recnorm}
\Tr_{2k-1} [ R^{(k)}] = I_{2k-2} \otimes R^{(k-1)} \qquad k=1, \dots, N~
\end{equation} 
where $R^{(k)} \in \Lin{\bigotimes_{j=0}^{2k-1} \spc H_j} $, and $R^{(0)} =1$.  Eq. (\ref{recnorm})
is the translation in terms of Choi operators of the fact that the $N$-partite channel $\map R^{(N)}
= \map C_{N-1} \circ \map C_{N-2} \circ \dots \circ \map C_0$, sending states on the even Hilbert
spaces $\St{\bigotimes_{k=0}^{N-1} \spc H_{2k}}$ to states on the odd ones
$\St{\bigotimes_{k=0}^{N-1} \spc H_{2k+1}}$, is a deterministic \emph{causal automaton}
\cite{wk,semidirk}, namely a channel where the reduced dynamics of an input state at step $k$
depends only on input states at steps $k'\le k$, and not at steps $k'> k$.  With different
motivations from supermaps and causal automata, Eq.  (\ref{recnorm}) also appeared in the work by
Gutoski and Watrous toward a general formulation of quantum games \cite{watgut}.

We call $\Detcomb{\bigotimes_{j=0}^{2N-1} \spc H_j}$ the set of
non-negative operators satisfying Eq.(\ref{recnorm}), and
$\Probcomb{\bigotimes_{j=0}^{2N-1} \spc H_j}$ the set
\begin{equation}
\Probcomb{\bigotimes_{j=0}^{2N-1} \spc H_j}=\left\{ R^{(N)} \ge 0 ~|~ \exists S^{(N)} \in \Detcomb{\bigotimes_{j=0}^{2N-1} \spc H_j}~:  R^{(N)} \le S^{(N)}  \right\}~.
\end{equation}  It is possible to prove that any operator $R^{(N)} \in \Detcomb{\bigotimes_{j=0}^{2N-1} \spc H_j}$  is the Choi operator of some sequential network of $N$ channels, or, equivalently, of some causal channel $\map R^{(N)}$ \cite{wk,comblong}. The minimal Stinespring dilation of the channel in terms of the Choi operator is given by \cite{covinst}
\begin{equation}\label{stine}
\map R^{(N)} (\rho) = \Tr_A[ V \rho V^\dag] \qquad V = \left(I_{odd} \otimes \sqrt{R^{(N) \tau}}\right) (|I_{odd}\kk \otimes I_{even})~,
\end{equation}
where $\spc H_{odd} = \bigotimes_{k=0}^{N-1} \spc H_{2k+1}$, $\spc H_{even} = \bigotimes_{k=0}^{N-1} \spc H_{2k}$,  $\tau$ denotes transposition w.r.t. a fixed orthonormal basis, $\spc H_{A} = \Supp \left({R^{(N)\tau} }\right)$ is the minimal ancilla space,  $|I_{odd}\kk$ is the unnormalized maximally entangled state on $\spc H_{odd}^{\otimes 2}$, and $V$ is an isometry from  $\spc H_{even}$ to $\spc H_{odd} \otimes \spc H_A$.

\subsection{Quantum $N$-instruments}
\begin{figure}
\setlength{\unitlength}{.7cm}
\begin{picture}(19,5.5)(0,0) 
  \put(0,2.75){\line(1,0){1}}
  \put(3,2.75){\line(1,0){.5}}
  \put(5.5,2.75){\line(1,0){.5}}
  \put(8,2.75){\line(1,0){.5}}
  \put(10,2.75){\line(1,0){.5}}
  \put(12.5,2.75){\line(1,0){.5}}
  \put(15,2.75){\line(1,0){.5}}
  \put(17.5,2.75){\line(1,0){1}}
  \put(1,0.5){\line(0,1){3}}
  \put(3,0.5){\line(0,1){3}}
  \put(1,0.5){\line(1,0){2}}
  \put(1,3.5){\line(1,0){2}}
  \put(6,0.5){\line(0,1){3}}
  \put(8,0.5){\line(0,1){3}}
  \put(6,0.5){\line(1,0){2}}
  \put(6,3.5){\line(1,0){2}}
  \put(15.5,0.5){\line(0,1){3}}
  \put(17.5,0.5){\line(0,1){3}}
  \put(15.5,0.5){\line(1,0){2}}
  \put(15.5,3.5){\line(1,0){2}}
  \put(10.5,.5){\line(0,1){3}}
  \put(12.5,.5){\line(0,1){3}}
  \put(10.5,.5){\line(1,0){2}}
  \put(10.5,3.5){\line(1,0){2}}
  \put(3,1.25){\line(1,0){3}}
  \put(8,1.25){\line(1,0){.5}}
  \put(10,1.25){\line(1,0){.5}}
  \put(12.5,1.25){\line(1,0){3}}
  \put(17.5,1.25){\line(1,0){.5}}
  \put(18,1.25){\oval(2.2,1.5)[r]}
  \put(18,.5){\line(0,1){1.5}} 

  \put(1.6,1.8){$\map C_0$}
  \put(6.5,1.8){$\map C_1$}
  \put(10.9,1.8){$\map C_{N-2}$}
  \put(15.9,1.8){$\map C_{N-1}$}
 
 \put(18.2,1){$P_B$}
  \multiput(9,2.75)(0.15,0){4}
  {\line(1,0){0.05}}
  \multiput(9,1.25)(0.15,0){4}
  {\line(1,0){0.05}}
\end{picture}
\caption{$N$-instrument: sequence of $N$  channels followed with postselection  on the last ancilla.   }\label{inst}
\end{figure}
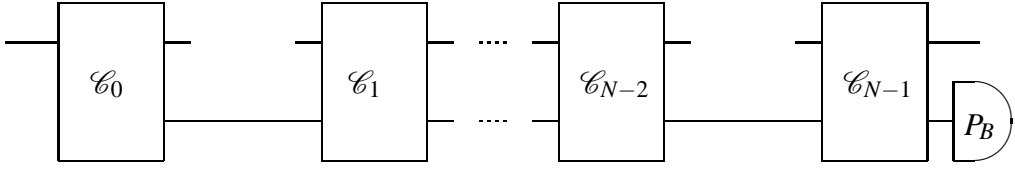

Let $\Omega$ be a measurable space and $\sigma (\Omega)$ be its
$\sigma$-algebra of events. A quantum $N$-instrument $R^{(N)}$ on
$\bigotimes_{j=0}^{2N-1} {\spc H_j}$ is an operator-valued measure
that associates to any event $B \in \sigma(\Omega)$ an $N$-comb
$R^{(N)}_B \in \Probcomb{\bigotimes_{j=0}^{2N-1} \spc H_j}$, and
satisfies the normalization
\begin{equation}
R^{(N)}_\Omega \in \Detcomb {\bigotimes_{j=0}^{2N-1} \spc H_j}~.
\end{equation} 
\begin{Theo}[Dilation of $N$-instruments]\label{reinst}
For any $N$-instrument $R^{(N)} $ on ${\bigotimes_{j=0}^{2N-1} \spc H_j}$ there exist a deterministic $N$-comb $S^{(N)} \in \Detcomb{\bigotimes_{j=0}^{2N-1} \spc H'_j}$ with $\spc H_j' = \spc H_j$ for $j=0, \dots, 2N-2$, and $\spc H_{2N-1}' = \spc H_{2N-1} \otimes \spc H_A$,   and a POVM $P$ on the ancilla $\spc H_A$  such that 
\begin{equation}
R^{(N)}_B = \Tr_{A}\left[   S^{(N)} ~ \left(  I_0 \otimes \dots \otimes I_{2N-1} \otimes P^\tau_B \right)  \right] \qquad \forall B \in \sigma(\Omega)~,  
\end{equation}  
 $\tau$ denoting transposition w.r.t. a fixed orthonormal basis.
\end{Theo}
The meaning of the theorem is that a quantum $N$-instrument can be always achieved by a
network of $N$ channels with postselection induced by the measurement
on an ancilla exiting from the $N$-th channel, as in Fig. \ref{inst}.

\Proof Diagonalize $R^{(N)}_\Omega$ as $R^{(N)}_\Omega = \sum_{i=1}^r \lambda_i
|\phi_i\>\<\phi_i|$, and take $\spc H_A = \Supp \left(R^{(N)\tau}_\Omega\right)=\Span\{|\phi^*_i\>~|~ i=1, \dots, r\}$,  $|\phi^*_i\> := \sum_n \<n|\phi_i\>^* |n\>$. 
Consider the purification $S^{(N)} = |R^{(N) \frac 1 2}_\Omega  \kk \bb
R^{(N) \frac 1 2}_\Omega|$ where $|R^{(N) \frac 1 2}_\Omega\kk = \sum_i
\sqrt{\lambda_i} |\phi_i\>|\phi^*_i\> \in \left( \bigotimes_{j=0}^{2N-1} \spc H_j \right) \otimes \spc H_A$.  By construction, $S^{(N)}$ is a
deterministic comb in $\Detcomb{\bigotimes_{j=0}^{2N-1} \spc H_j'}$.
Now define the POVM $P$ by $P_B = \left[R^{(N)}_\Omega\right]^{-\frac 1 2}~ R^{(N)}_B~ \left[R^{(N)}_\Omega \right]^{- \frac 1 2}$, where $\left[R^{(N)}_\Omega\right]^{-\frac 1 2}$ is the inverse of $R^{(N) \frac 1 2}_\Omega$ on its support. It is immediate to check that for any event $B \in \sigma (\Omega)$ we have $R^{(N)}_B = \Tr_{A} [S^{(N)} ~ (I_0 \otimes \dots \otimes I_{2N-1} \otimes P_B^\tau)] $. \qed   
This theorem is similar in spirit to Ozawa's dilation theorem for quantum instruments \cite{ozawa}. The important difference here that $P$ is a POVM on a finite-dimensional ancilla space, rather than a von Neumann measurement in infinite dimension. 
\subsection{Quantum $N$-testers} 
An $N$-tester $T^{(N)} $ is an $(N+1)$-instrument where the
first and last Hilbert spaces, $\spc H_0$ and $\spc H_{2N+1}$, respectively, are
one-dimensional. Accordingly, we can shift back by one unit the numeration of Hilbert spaces, so that, if  $B \in \sigma (\Omega)$ is an event, then  $T^{(N)}_B$ is an operator on $\bigotimes_{j=0}^{2N-1} \spc H_j$.  With this shifting, the normalization of the tester is given by
\begin{equation}\label{testnorm}
\begin{split}
&T^{(N)}_\Omega  = I_{2N-1} \otimes \Xi^{(N-1)} \\
&\Tr_{2k-2}[\Xi^{(k)}] = I_{2k-3} \otimes \Xi^{(k-1)} \qquad k = 2, \dots, N~\\
&\Tr_0[\Xi^{(1)}]=1 ~,     
\end{split}
\end{equation} 
with $\Xi^{(k)}  \in \Lin{\bigotimes_{j=0}^{2k-2} \spc H_j}$. 

\begin{figure}[h]
\setlength{\unitlength}{.7cm}
\begin{picture}(19,5.5)(0,0) 
 \put(3,2.75){\line(1,0){.5}}
  \put(5.5,2.75){\line(1,0){.5}}
  \put(8,2.75){\line(1,0){.5}}
  \put(10,2.75){\line(1,0){.5}}
  \put(12.5,2.75){\line(1,0){.5}}
  \put(15,2.75){\line(1,0){.5}}
  \put(3,2){\oval(4,3)[l]}
  \put(3,0.5){\line(0,1){3}}

  \put(6,0.5){\line(0,1){3}}
  \put(8,0.5){\line(0,1){3}}
  \put(6,0.5){\line(1,0){2}}
  \put(6,3.5){\line(1,0){2}}
  \put(3.5,2){\line(0,1){3}}
  \put(5.5,2){\line(0,1){3}}
  \put(3.5,2){\line(1,0){2}}
  \put(3.5,5){\line(1,0){2}}
  \put(13,2){\line(0,1){3}}
   \put(15,2){\line(0,1){3}}
  \put(13,2){\line(1,0){2}}
  \put(13,5){\line(1,0){2}}
  \put(10.5,.5){\line(0,1){3}}
  \put(12.5,.5){\line(0,1){3}}
  \put(10.5,.5){\line(1,0){2}}
  \put(10.5,3.5){\line(1,0){2}}
  \put(3,1.25){\line(1,0){3}}
  \put(8,1.25){\line(1,0){.5}}
  \put(10,1.25){\line(1,0){.5}}
  \put(12.5,1.25){\line(1,0){3}}
  \put(5.5,4.25){\line(1,0){2.5}}
   \put(10.5,4.25){\line(1,0){2.5}}
  \put(15.5,2){\oval(4.3,3)[r]}
   \put(15.5,0.5){\line(0,1){3}} 

  \put(1.6,1.8){$\rho_0$}
  \put(6.5,1.8){$\map D_1$}
  \put(10.8,1.8){$\map D_{N-1}$}
  \put(4.3,3.3){$\map C_0$}
  \put(13.4,3.3){$\map C_{N-1}$}
 \put(16.2,1.8){$P_B$}
  \multiput(9,2.75)(0.15,0){4}
  {\line(1,0){0.05}}
  \multiput(9,1.25)(0.15,0){4}
  {\line(1,0){0.05}}
  \multiput(9,4.25)(0.15,0){4}
  {\line(1,0){0.05}}
\end{picture}
\caption{\label{tester} Testing a network of $N$ quantum operations $(\map C_i)_{i=0}^{N-1}$. The $N$-tester consists in the preparation of an input state
  $\rho_0$,  followed by channels $\{\map D_1, \dots, \map D_{N-1}\}$, and a final
  measurement  $P_B$.}
\end{figure}
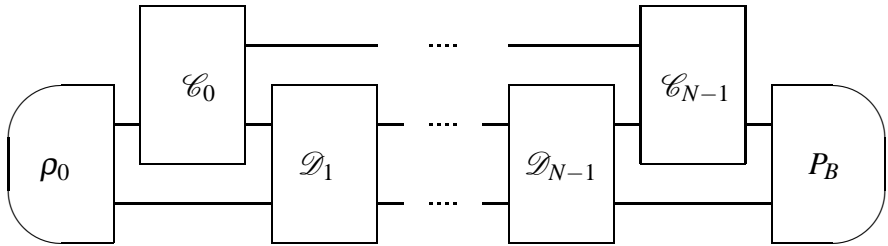

A tester represents a quantum network starting with a state
preparation and finishing with a measurement on the ancilla.  When
such a network is connected to a network of $N$ quantum operations as
in Fig. \ref{tester}, the only outputs are measurement outcomes.

Precisely, if the comb of the measured network is $R^{(N)} \in
\Probcomb{\bigotimes_{j=0}^{2N-1} \spc H_j}$, then the probabilities of
any event are given by the generalized Born rule
\cite{combs,ziman}
\begin{equation}\label{Born}
p(B|R^{(N)}) = \Tr[T^{(N)\tau}_B ~ R^{(N)} ] \qquad \forall B\in \sigma (\Omega)~. 
\end{equation}  
For deterministic combs  $R^{(N)} \in
\Detcomb{\bigotimes_{j=0}^{2N-1} \spc H_j}$  the  probabilities sum up to one: 
\begin{equation}\label{normprob}
p(\Omega| R^{(N)})  = \Tr[T^{(N)\tau}_\Omega R^{(N)}] =1~.
\end{equation}
Clearly, since $T^{(N)\tau} $ is also a tester, the Born rule
can be written in the familiar way without the transpose.
However, here we preferred to write probabilities in terms of the
combs $R^{(N)}$ and $T^{(N)}_B$ of the measured and measuring
networks, respectively. In fact, the Born rule is nothing but a
particular case of link product \cite{combs}, and the transpose appears as the signature of the interlinking of the two
networks.

\begin{Prop}[Decomposition of $N$-testers \cite{memorydisc}]\label{retest}
  Let $T^{(N)} $ be a quantum $N$-tester on
  $\bigotimes_{j=0}^{2N-1} \spc H_j$, and consider the ancilla space $\spc H_A = \Supp \left( T^{(N)\tau}_\Omega \right)$.  
Let $\map S$ be the linear supermap from
  $\Probcomb{\bigotimes_{j=0}^{2N-1} \spc H_j}$ to
   $\St{ \spc H_A}$ given by
\begin{equation}
\map S(R^{(N)}) =\left[ T^{(N) \tau}_\Omega \right]^{\frac 1 2} ~R^{(N)}~  \left[T^{(N)  \tau}_\Omega \right]^{\frac 1 2}   
\end{equation}
and $P$ be the POVM on $\spc H_A$ defined by 
\begin{equation}
P_B = \left[T^{(N)}_\Omega \right]^{-\frac 1 2} ~ T^{(N)}_B ~\left[ T^{(N)}_\Omega \right]^{-\frac 1 2}~.
\end{equation}
The supermap $\map S$ transforms deterministic combs into normalized states of the ancilla. The probabilities of events are given  by
\begin{equation}\label{testerdeco}
p(B|R^{(N)}) = \Tr[T^{(N)\tau}_B ~ R^{(N)}] = \Tr [P_B^\tau \map S(R^{(N)})]~. 
\end{equation}  
\end{Prop}
 This proposition reduces any measurement on an input quantum network to a measurement on a suitable state, which is obtained by linear transformation of the input comb. As we will see in the following, this simple result has very strong consequences in quantum estimation.     
\Proof If $R^{(N)}$ is in $\Detcomb{\bigotimes_{j=0}^{2N-1} \spc H_j}$, then $\Tr[\map S(R^{(N)})] = \Tr[T^{(N) \tau}_\Omega R^{(N)}] =1$, having used Eq. (\ref{normprob}).  Eq. (\ref{testerdeco}) is an obvious consequence of the definitions of $\map S$ and $P$. \qed 

Proposition \ref{retest} reduces the discrimination of two networks to the discrimination of two states. This allows us to define an operational notion of distance between networks \cite{memorydisc}, whose meaning is directly related to minimum error discrimination:
\begin{equation}
  \left|\! \left | R^{(N)} - R^{(N)'}  \right| \!\right|_{op}  = \max_{T^{(N)}_\Omega  }  \left| \! \left | \left[ T^{(N)  \tau}_\Omega \right]^{\frac 1 2} \left(  R^{(N)} - R^{(N) '}  \right) \left[T^{(N)  \tau}_\Omega \right]^{\frac 1 2} \right|\!\right|_1~,   
\end{equation}  
with $|\!| A |\!|_1 = \Tr |A|$.  Remarkably, the above norm can be
strictly greater than the diamond (cb) norm of the difference $\map
R^{(N)} - \map R^{(N)'}$ of the two multipartite channels \cite{memorydisc}.
This means that a scheme such as in Fig. \ref{tester} can achieve
a strictly better discrimination than a parallel scheme where a multipartite entangled state is fed in the unknown channel and a multipartite measurement is performed on the output. 
\section{Covariant quantum networks}
\subsection{Covariant $N$-combs}  
Let $\grp G$ be a group, acting on the Hilbert space $(\spc H_j)_{j=0}^{2N-1}$
via the a unitary representation $\{U_{g,j}~|~ g \in \grp G\}$. Denote by $\map U_{g,j}$ the map $\map U_{g,j} (\rho)=U_{g,j} \rho U_{g,j}^\dag$.  Suppose that the causal channel $\map R^{(N)}$ from $\St{\bigotimes_{k=0}^{N-1}  \spc H_{2k}}$ to $\St{\bigotimes_{k=0}^{N-1} \spc H_{2k+1}}$ is \emph{covariant}, namely 
\begin{equation}
\map R^{(N)} \circ  \left(\bigotimes_{k=0}^{N-1}  \map U_{g,2k} \right) (\rho) = \left(\bigotimes_{k=0}^{N-1} \map U_{g,2k+1}\right) \circ \map R^{(N)} (\rho)~. 
\end{equation}
Then the corresponding comb, which we call \emph{covariant} either, satisfies the commutation property
\begin{equation}
\left[R^{(N)},\bigotimes_{k=0}^{N-1} (U_{g, 2k+1} \otimes U^*_{g,2k}) \right]=0 \qquad \forall g \in \grp G~.\end{equation}
For covariant combs, the minimal dilation of the memory channel $\map R^{(N)}$ given by Eq. (\ref{stine}) satisfies the commutation relation
\begin{equation}
\left[ \left(\bigotimes_{k=0}^{N-1} U_{g, 2k+1} \right) \otimes U_{g,A} \right] V = V \left(\bigotimes_{k=0}^{N-1} U_{g,2k}\right) ~,
\end{equation}
where $U_{g,A}$ is the compression of $\left(\bigotimes_{k=0}^{N-1} (U^*_{g, 2k+1} \otimes U_{g,2k}) \right)$ to the invariant subspace $\spc H_A = \Supp\left( R^{(N)\tau}\right)$. 

\subsection{Covariant $N$-instruments and testers}   
Suppose that the group $\grp G$ acts on the outcome space $\Omega$. For $B \in \sigma (\Omega)$, denote by $gB :=\{g \omega~|~ \omega \in B\}$. A covariant $N$-instrument  $R^{(N)} $ is defined by the property
\begin{equation}\label{covinst}
  R^{(N)}_{g B} = \left( \bigotimes_{k=0}^{N-1} (\map U_{g, 2k+1} \otimes \map U^*_{g,2k})\right) \left( R^{(N)}_B\right)~.
\end{equation} 
A covariant tester is simply a covariant $N$-instrument with one-dimensional $\spc H_0$ and $\spc H_{2N-1}$ and with all remaining labels shifted back by unit. We now suppose that $\grp G$ is compact and $\Omega$ is transitive, i.e. for any pair $\omega_1, \omega_2 \in \Omega$ there always exists a group element $g\in \grp G$ such that $\omega_2 = g \omega_1$.
\begin{Theo}[Structure of covariant $N$-instruments/testers] 
  Let $\grp G$ be compact and $\Omega$ be transitive, with normalized
  Haar measure $\d \omega$. Let $\omega_0 \in \Omega$ be a point of
  $\Omega$, 
  and let $\grp G_0=\{g\in \grp G~|~ g \omega_0 = \omega_0 \}$ be the
  stabilizer of $\omega_0$. Let $\sigma: \Omega \to \grp G$ be a
  measurable section, such that $\omega = \sigma_\omega \omega_0$.  If
  $R^{(N)} $ is a covariant instrument, then there exists a
  non-negative operator $D^{(N)}_0$ such that
\begin{equation}
\begin{split}
  &R^{(N)}_B  = \int_B \d \omega ~  D^{(N)}_\omega \\
& D^{(N)}_\omega = \left (\bigotimes_{k=0}^{N-1} ( U_{\sigma_{\omega}, 2k+1} \otimes  U^*_{\sigma_{\omega},2k}) \right) ~ D^{(N)}_0~ \left(\bigotimes_{k=0}^{N-1} ( U_{\sigma_{\omega}, 2k+1} \otimes U^*_{\sigma_{\omega},2k}) \right)^\dag\\
  &\left [D_0^{(N)}, \bigotimes_{k=0}^{N-1} (U_{g_0,2k+1} \otimes U_{g_0,2k}^*)\right] =
  0 \qquad \forall g_0 \in \grp G_0~.
\end{split}
\end{equation}  
\end{Theo} 
\Proof Simple generalization of the standard proof for covariant POVMs \cite{HolevoBook}.

For a covariant $N$-instrument/tester, Eq. (\ref{covinst}) implies the commutation
\begin{equation}
\left[R^{(N)}_\Omega, \bigotimes_{k=0}^{N-1} (U_{g,2k+1} \otimes U_{g,2k}^*) \right]=0 \qquad \forall g \in \grp G~.
\end{equation}
This implies additional group structure in the results of Theorem
\ref{reinst} and Proposition \ref{retest}. In particular, for
covariant testers, the map $\map S: \Probcomb{\bigotimes_{j=0}^{2N-1}
  \spc H_j} \to \St{\spc H_A}$ is a \emph{covariant supermap}:
\begin{equation}\label{usami}
\map S \circ \left( \bigotimes_{k=0}^{N-1} (\map U^*_{g,2k+1} \otimes \map U_{g,2k})  \right)  \left(R^{(N)} \right) = U_{g, A} ~\map S(R^{(N)}) ~ U_{g,A}^\dag~, 
\end{equation} 
where $U_{g,A}$ is the compression of $\bigotimes_{k=0}^{N-1} \left( U^*_{g,2k+1} \otimes U_{g, 2k}\right) $ to the invariant subspace $\spc H_A = \Supp \left( T^{(N)\tau}_\Omega \right)\subseteq \bigotimes_{j=0}^{2N-1} \spc H_j$.
\section{Optimal covariant estimation of quantum networks}
Let $\left \{R^{(N)}_{\omega} \in \Detcomb{\bigotimes_{j=0}^{2N-1} \spc
  H_j}~|~ \omega \in \Omega\right\}$ be a family of quantum networks
parametrized by $\omega$. We now want to find the optimal tester to
estimate the parameter $\omega$.  For simplicity, we consider here the
special case in which $\Omega \equiv \grp G$, for some compact group $\grp G$, and $R^{(N)}_g$ has the form 
\begin{equation}
R^{(N)}_g =     \left( \bigotimes_{k=0}^{N-1} (U_{g,2k+1} \otimes U_{g,2k}^*) \right) ~ R^{(N)}_0~  \left( \bigotimes_{k=0}^{N-1} (U_{g,2k+1} \otimes U_{g,2k}^*)\right)^\dag
\end{equation}  Let $c(\hat g, g)$ be a cost
function, penalizing the differences between the estimated parameter
$\hat g$ and the true one $g$. Suppose that $c(\hat g,g)$ is
left-invariant, namely $c(h\hat g , hg) = c(\hat g, g)~ \forall h \in
\grp G$. The optimal estimation is then given by the tester $T^{(N)}$ that minimizes the average cost
\begin{equation}
\<c\> = \int_{\grp G} \d g ~ \int_{\grp G} ~ c(\hat g, g) ~ \Tr[T^{(N)\tau}_{\d \hat g}  R^{(N)}_g]~,  
\end{equation}  
where $\d g$ is the normalized Haar measure, and $\int_{\grp G} f(\hat
g)~ \Tr\left[ T^{(N)\tau}_{\d \hat g} R^{(N)} \right]$ denotes integration
of $f$ against the scalar measure $\mu_B = \Tr[T^{(N)\tau}_B R^{(N)} ]$.
An alternative notion of optimality is the minimization of the
worst-case cost
\begin{equation}
c_{wc} =\max_{g \in \grp G}   \left(\int_{\grp G} c(\hat g, g) ~ \Tr[T^{(N)\tau}_{\d \hat g} R^{(N)}_g ]  \right)~.
\end{equation}
However, it it easy to prove that in the covariant setting it is sufficient to consider covariant testers, for which the average and worst-case cost coincide:
\begin{Theo}
There exists a covariant tester $T^{(N)}_B = \int_B \d g~ D^{(N)}_g$, with density 
\begin{equation}
D^{(N)}_g = \left( \bigotimes_{k=0}^{N-1} (U^*_{g,2k+1} \otimes U_{g,2k})\right) D^{(N)}_0 \left(\bigotimes_{k=0}^{N-1} (U_{g,2k+1}^* \otimes U_{g,2k}) \right)^\dag 
\end{equation}
 that is optimal both for the average and worst-case cost.     
\end{Theo}
\Proof The standard averaging argument \cite{HolevoBook}: if $T^{(N)} $ is an
optimal tester, then the tester $\overline{T^{(N)}}$ defined by
$\overline{T_B^{(N)}} = \int_{\grp G} \d h \bigotimes_{k=0}^{N-1}
(\map U_{h,2k+1}^* \otimes \map U_{h,2k}) \left(T^{(N)}_{h^{-1}B}\right)$ is covariant
and has the same average and worst-case cost as $T^{(N)} $.
Moreover, for any covariant tester, the average and worst-case cost
coincide.

\section{Applications}
\subsection{Optimal estimation of group transformations with $N$
  copies} Suppose we have at disposal $N$ uses of a black box
performing the unknown group transformation $U_g$, and that we want to
find the optimal network for estimating $g$. In this case the
parametric family of networks is $R^{(N)}_g = (|U_g\kk\bb
U_g|)^{\otimes N}$, where $|U_g\kk := (U_g \otimes I) |I\kk$, $|I\kk =
\sum_{i=1}^d |i\>|i\> $. Using Proposition \ref{retest} and Eq.
(\ref{usami}), the optimal estimation on these networks is reduced to
the optimal estimation on the ancilla states $\rho_g = \map S \left(
R_g^{(N)}\right) = U_{g,A} \map S \left(R^{(N)}_0\right) U_{g, A}^\dag$, with $R^{(N)}_0 =
(|I\kk\bb I|)^{\otimes N}$. Since the ancilla space is an invariant
subspace of $\bigotimes_{j=0}^{2N-1} \spc H_j$ and the representation
$U_{g,A}$ is a sub-representation of $\bigotimes_{k=0} (U_{g, 2k+1}
\otimes I_{2k})$, it is clear that the minimum cost in the estimation
is lower bounded by the minimum cost achievable in a parallel scheme,
where the unitary $U_g^{\otimes N}\otimes I_{ref}$ is applied to a
multipartite entangled state in $\St{ \bigotimes_{k=1}^N \spc H_k
  \otimes \spc H_{ref}}$, with  $\spc
H_{ref}$  suitable reference space. In this way the optimal estimation is reduced to the
 optimal parallel estimation of Ref. \cite{EntEstimation}.

\subsection{Optimal alignment of reference frames with multi-round protocols}
Two distant parties Alice and Bob, who lack a shared reference frame,
can try to establish one by sending suitable physical systems, such as
clocks and gyroscopes for time and orientation references,
respectively.  In the quantum scenario, the role of elementary clocks
and gyroscopes is played by spin $1/2$ particles, and it has been
shown that the optimal protocol using $N$ particles in a single round of quantum
communication from Alice to Bob has a r.m.s. error scaling to zero as
$1/N$ (with suitable constants) for both for clock
synchronization \cite{optclocks} and Cartesian axes alignment \cite{refframe}. However, the optimal protocol for
establishing reference frames with many rounds of quantum
communication and arbitrary amount of classical
communication has been not analyzed yet. In principle, an adaptive strategy might improve the alignment, if not by changing the scaling with $N$, at least by improving the constant. With the formalism of covariant combs and testers, however, it is rather straightforward to prove that this is not the case.  

Let us consider the general case in which the mismatch between Alice's
and Bob's reference frames is represented by an unknown element $g$ of
some group of physical transformations $\grp G$. The unitary
(projective) representation in the Hilbert spaces of quantum systems
yields the passive transformation of states due to the change from
Alice's to Bob's viewpoint: a single-particle state that is $|\psi^{(A)}\>$ is
Alice's reference frame becomes $|\psi^{(B)}\>= U_g|\psi^{(A)}\>$ in
Bob's one, a single-particle operator $O^{(A)}$ becomes $O^{(B)} = U_g O^{(A)}
U_g^\dag$, and a single-particle operation $\map C^{(A)}$ becomes $\map C^{(B)} =\map U_g \map C^{(A)} \map U_g^\dag$. Consider a protocol with $2r$ rounds of quantum
communication ($r$ rounds from Alice to Bob and $r$ from Bob to Alice)
with $q_i$   quantum particles exchanged per round.  We also allow an
unbounded amount of classical communication, represented by the
exchange of $\grp G$-invariant systems prepared in classical
(diagonal) states.  The goal of the protocol is to give the best
possible estimate $\hat g$ of the mismatch $g$.  Notice that, since
Alice and Bob are not restricted in sending classical data, we can
imagine without loss of generality that the estimate $\hat g$ is
produced by Bob (if it were produced by Alice, she could always
transmit this classical information to Bob). The protocol is then
represented by the interlinking of two networks of quantum operations:
\emph{i)} Alice's network is a deterministic $r$-comb $R^{(r,A)} \in
\Detcomb{\spc H_{A \to B} \otimes \spc H_{B \to A} \otimes \spc H_C
}$, where $\spc H_{A \to B}$ ($\spc H_{B\to A}$) is the Hilbert space
of all particles sent from Alice to Bob (from Bob to Alice), and $\spc
H_C$ is the Hilbert space of the invariant systems used for classical
communication, and $\emph{ii)}$ Bob's network is an $r$-tester
$T^{(r,B)}_{\d \hat g}$ on the same Hilbert spaces.  When switching to
Bob's reference frame, all Alice's operations are conjugated by
unitaries, and her comb becomes
\begin{equation}
R^{(r, B)}_g = \left( U_g^{\otimes N_{A \to B}} \otimes U_g^{* \otimes N_{B\to A}} \otimes I_C \right) ~  R^{(r,A)} ~ \left( U_g^{\otimes N_{A \to B}} \otimes U_g^{* \otimes N_{B\to A}} \otimes I_C \right)~.  
\end{equation}
where $N_{A \to B}$ ($N_{B \to A}$) is the number of particles
traveling from Alice to Bob (from Bob to Alice). Notice that we have
the identity $I_C$ on the classical systems, since classical
communication (strings of bits) is invariant under changes of
reference frame.  Therefore, for any left-invariant cost function
$c(\hat g, g)$ we are in the case of covariant network estimation
treated before. The estimation of $g$ from the networks $R^{(r,B)}_g$
is then reduced to the estimation of $g$ from the states $\rho_g =
\map S (R^{(r,B)}_g) = U_{g,A} \rho_0 U_{g,A}^\dag$, where $U_{g,A}$
is a sub-representation of $ U_g^{\otimes N_{A \to B}} \otimes U_g^{*
  \otimes N_{B\to A}} \otimes I_C$.  For $\grp G = U(1)$ and $\grp G =
\SU 2$ $U_g$ and $U_g^*$ are equivalent representations (up to global phases), hence this is
exactly the same estimation that can be achieved by sending $N_{A\to
  B} + N_{B \to A}$ particles in a single round. Even for groups for
which $U_g$ and $U_g^*$ are not equivalent (such as $\SU d$), one can
achieve the same estimation precision in a single round by sending
$N_{A \to B}$ particles and $N_{B\to A}$ charge-conjugate particles
from Alice to Bob.  This proves that anyway there is no advantage in
using more than one round of quantum communication, and that classical
communication is completely useless.

\bigskip

This work is supported by the EC through the project CORNER. G. C. acknowledges financial support
from the Risk and Security Study Center, IUSS Pavia.

\end{document}